\documentclass[aps,prl,twocolumn,reprint,superscriptaddress,citeautoscript,longbibliography]{revtex4-1}

\usepackage[dvipdfmx]{graphicx}
\usepackage{amsmath,amssymb}
\usepackage{fixmath}
\usepackage[normalem]{ulem}

\usepackage{color}

\begin{document}

\title{Einstein--de Haas Nanorotor}

\author{W. Izumida}
\email[]{wizumida@tohoku.ac.jp}
\affiliation{Department of Physics, Tohoku University, Sendai 980-8578, Japan}
\author{R. Okuyama}
\affiliation{Department of Physics, Meiji University, Kawasaki 214-8571, Japan}
\author{K. Sato}
\affiliation{National Institute of Technology, Sendai College, Sendai 989-3128, Japan}
\author{T. Kato}
\affiliation{Institute for Solid State Physics, University of Tokyo, Kashiwa 277-8581, Japan}
\author{M. Matsuo}
\affiliation{Kavli Institute for Theoretical Sciences, University of Chinese Academy of Sciences, Beijing, 100190, China.}
\affiliation{CAS Center for Excellence in Topological Quantum Computation, University of Chinese Academy of Sciences, Beijing 100190, China}
\affiliation{Advanced Science Research Center, Japan Atomic Energy Agency, Tokai, 319-1195, Japan}
\affiliation{RIKEN Center for Emergent Matter Science (CEMS), Wako, Saitama 351-0198, Japan}

\begin{abstract}

We propose a nanoscale rotor embedded between two ferromagnetic electrodes that is driven by spin injection. The spin-rotation coupling allows this nanorotor to continuously receive angular momentum from an injected spin under steady current flow between ferromagnetic electrodes in an antiparallel magnetization configuration.  We develop a quantum theory of this angular-momentum transfer and show that a relaxation process from a precession state into a sleeping top state is crucial for the efficient driving of the nanorotor by solving the master equation.  Our work clarifies a general strategy for efficient driving of a nanorotor.
\end{abstract}

\maketitle

{\it Introduction.---}
The angular-momentum conversion phenomena between spin and mechanical rotation are recognized as the gyromagnetic effects discovered in the early 20th century by S. J. Barnett, A. Einstein, and W. J. de Haas. They observed magnetization induced by mechanical rotation~\cite{Barnett1915} and mechanical rotation induced by magnetization~\cite{Einstein_deHaas_1915} known as the Einstein-de Haas (EdH) effect, revealing the origin of magnetism is the angular momentum.
While the original targets were limited to ferromagnets~\cite{Scott1962}, the gyromagnetic effects are universal phenomena that appeared even in nonmagnetic materials ranging from macroscopic to microscopic scales in various branches of physics, including ultra-cold atoms~\cite{Kawaguchi2006,Gawryluk2007}, spintronics~\cite{Takahashi2016,Kobayashi2017,Kurimune2020,Takahashi2020,Kazerooni2020,Kazerooni2021}, nuclear spin physics~\cite{Chudo2014}, and quark-hadron physics~\cite{Adamczyk2017}. Recently, the gyromagnetic effect is used to identify the ultrafast demagnetization process~\cite{Dornes2019}.
The gyromagnetic effect has also been utilized to generate mechanical torsion using electron spin relaxation on nanoscale objects~\cite{Mohanty-PRB-2004,Wallis-APL-2006-09,Zolfagharkhani-NatNano-2008,Harii-2019-NatCommun}, showing its potential as a driving force for mechanical vibrations in nanoelectromechanical systems~\cite{Cleland-2003-book}.

Recently, mechanical rotational motion has became a topic of nanoscale research
~\cite{Fennimore-Nature-2003,Bailey-PRL-2008}. 
For instance, the inner carbon nanotubes 
of multi-wall carbon nanotubes~\cite{Iijima-1991}, or encapsulated fullerenes in carbon peapods~\cite{Smith-nature-1998} may work as nanorotors while the outer nanotubes may act as the frame since the inner carbons are low friction due to the van der Waals bonded frame~\cite{Cumings-Zettl-2000,Cook2013,Huang2018}.
However, efficient driving forces for the rotational motion are still lacking~\cite{Fennimore-Nature-2003,Bailey-PRL-2008,footnote1}.

\begin{figure}[tb]
  \includegraphics[width=7.5cm]{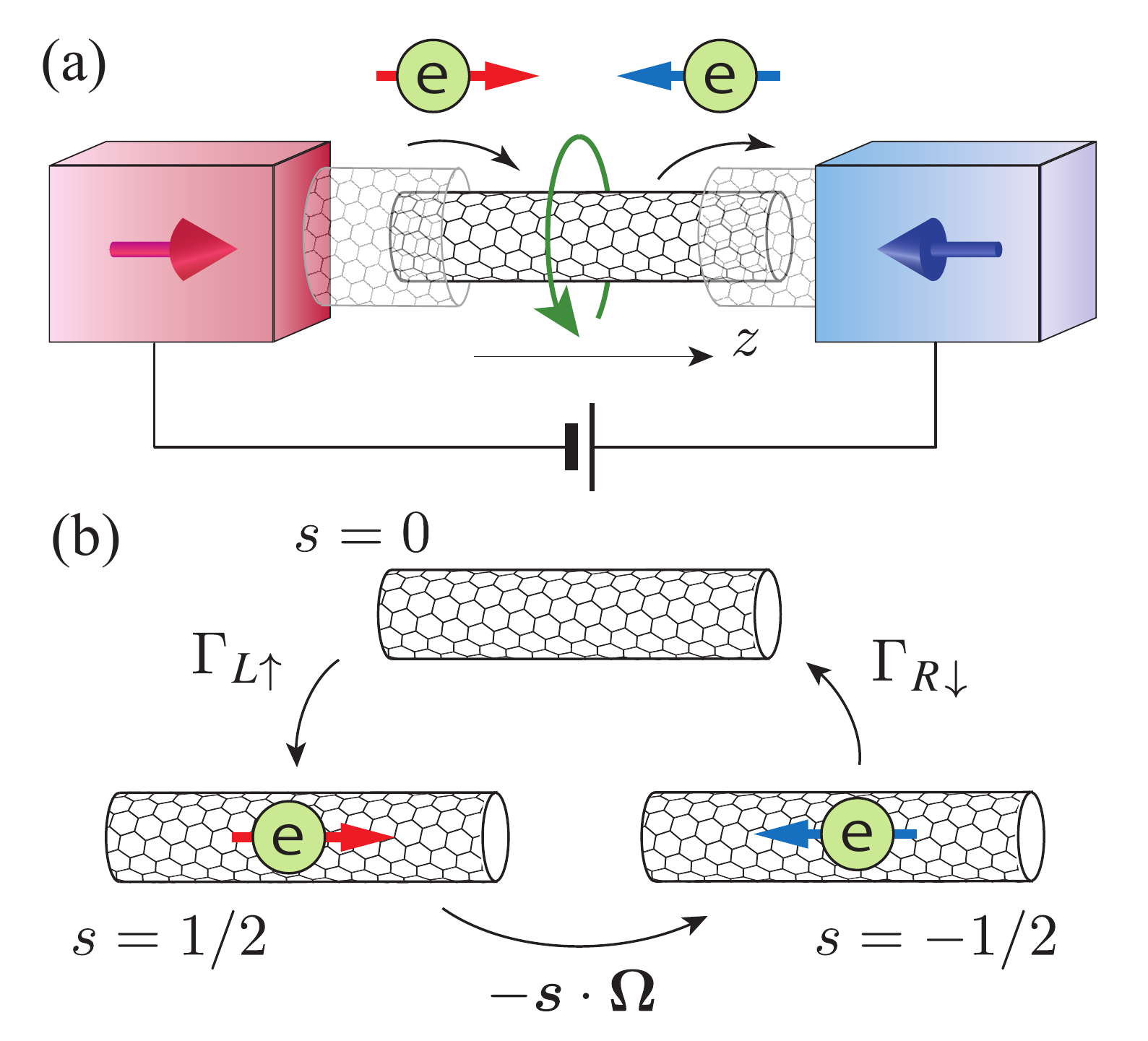}
  \caption{(a) Schematic diagram of proposed structure of nanorotor driven by spin injection.
    A voltage is applied to two half-metallic ferromagnetic electrodes with magnetizations respectively in the $z$ and $-z$ directions.
    An electron with a spin $s_z = \hbar/2$ 
    from the left electrode cannot enter the right electrode because there is no electronic state for a spin $s_z =  \hbar/2$.
    Continuous current flow is allowed via spin flipping at a rotor induced by the spin-rotation coupling that describes angular-momentum transfer from a spin to the rotor.
    (b) Three possible electronic states of the rotor.
    The number of additional electrons injected into the rotor is restricted to 0 or 1 because of the Coulomb blockade.
    The spin-rotation coupling is described by the Hamiltonian $- \mathbold{s} \cdot \mathbf{\Omega}$, and $\Gamma_{L\uparrow}$($\Gamma_{R\downarrow}$) is an electron transfer rate between the rotor and the left(right) electrode. }
  \label{fig:geometry}
\end{figure}

In this Letter, we develop a quantum theory to describe the microscopic mechanism of a nanorotor driven by electron spin injection via the EdH effect. 
As a feasible setup, we consider the double-wall carbon nanotube rotor shown in Fig.~\ref{fig:geometry}(a).
Such a rotor is experimentally feasible in ferromagnetic-carbon nanotube hybrid systems~\cite{Cottet-2006-10}.
To induce its rotational motion by electron spin injection, the nanorotor is embedded between two half-metallic ferromagnetic electrodes in an antiparallel magnetization configuration where the source and drain electrodes are oriented to $z$ and $-z$ directions, respectively.
The rotor is driven by cyclic transitions among its three electronic states, as shown in Fig.~\ref{fig:geometry}(b).
By applying a bias voltage between the electrodes, an electron with a spin $s_z = \hbar/2$ tunnels from the left electrode into the nanorotor.
This electron cannot tunnel to the drain electrode as long as its spin state remains $s_z = \hbar/2$ because there is no electronic state for the spin $s_z = \hbar/2$ in the half-metallic ferromagnetic drain.
Electron accumulation by additional electron tunneling is also forbidden because of the Coulomb blockade in the nanorotor that acts like a quantum dot~\cite{RevModPhys.87.703,Izumida-2015-06,Izumida-2016-05}.
Therefore, only spin flipping of the injected electron enables continuous current flow through the rotor.
Simultaneously with the single spin flipping ($s_z=\hbar/2 \rightarrow-\hbar/2$), angular momentum $\hbar$ is eventually transferred from the injected electron to the mechanical rotational motion.
Therefore, the current through the nanorotor is expected to be a driving force of the rotational motion of the nanorotor in this system.
To confirm the feasibility of the above idea, a concrete theory for the microscopic mechanism of angular-momentum transfer in the proposed setup is the subject in this Letter.
The obtained conditions for efficient driving are expected to be common to other experimental realizations.

\begin{figure}[tb]
  \includegraphics[width=8.5cm]{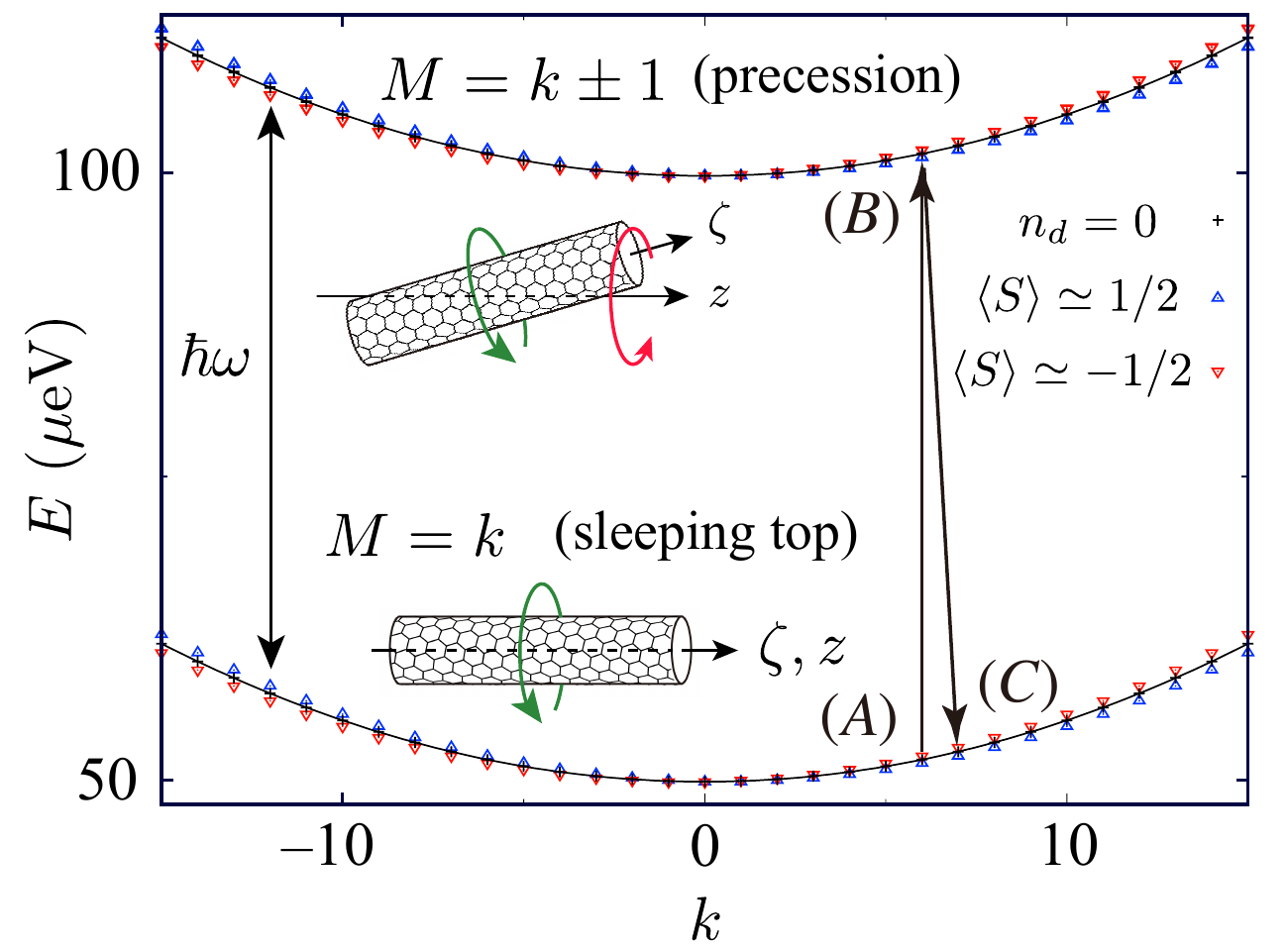}
  \caption{Eigenenergies of rotational motion as a function of $k$.
    The upper and lower branches correspond to the precession state ($M= k\pm 1$) and the sleeping top state ($M=k$), respectively.
    The insets are schematic diagrams of the corresponding rotor motions.
}
  \label{fig:en-k}
\end{figure}

We shall treat the mechanical rotation of the nanorotor as that of a 
rigid body~\footnote{Such a treatment is justified for nanoscaled objects such as a carbon nanotubes in which the typical energy scale of lattice deformation ($\sim$ meV) is much higher than that of the rotational motion.}.
In the present rotor, two kinds of rigid-body motion, i.e., precession and sleeping top as observed in classical tops, are possible at low energies (see the inset of Fig.~\ref{fig:en-k}).
Here we describe how to induce the sleeping top state, a rotational motion of the nanorotor around a principal axis ($\zeta$) with the axis aligned with an outer axis ($z$).
The process of angular-momentum transfer is modeled by the spin-rotation (SR) coupling~\cite{Hehl1990,Matsuo-JPSJ-2017}
\begin{equation}
  H_\mathrm{SR} = - \mathbold{s} \cdot \mathbf{\Omega},
  \label{eq:spin-rotor-int}
\end{equation}
which expresses how the electron spin $\mathbold{s}$ interacts with an effective magnetic field induced by a mechanical rotation with angular velocity $\mathbf{\Omega}$.
As discussed later, a relaxation mechanism for rotational motion in which angular momentum is exchanged with a bosonic bath can be incorporated to stabilize the rotational motion to the sleeping top state.

{\it Model.---} 
Let $(\xi,\eta,\zeta)$ be the principal axes of inertia and $(x,y,z)$ be the laboratory coordinates.
The moments of inertia of the corresponding principal axes are denoted as $I_{\xi}$, $I_{\eta}$, and $I_{\zeta}$, respectively.
We assume that the rotor is a symmetric top with $I_\xi = I_\eta$.
The rotational motion of the rotor is expressed by the following Hamiltonian~\cite{Landau-QM-1977},
\begin{equation}
  H_{R}
  = \frac{1}{2 I_\xi} \mathbold{L}^2 
  + \frac{1}{2} \left( \frac{1}{I_\zeta} - \frac{1}{I_\xi} \right) L_\zeta^2,
  \label{eq:HR}
\end{equation}
where $\mathbold{L}$ is the angular-momentum operator of the rotor, and $L_\zeta$ is the $\zeta$ component of $\mathbold{L}$.
The Hamiltonian~\eqref{eq:HR} simply expresses free rotational motion.
The quantum states of the rotor, i.e., the eigenstates of Eq.~\eqref{eq:HR}, are denoted as $| L, M, k \rangle$, which is specified by the three quantum numbers, the amplitude of angular momentum $L$, its $z$ component $M$, and the $\zeta$ component $k$~\cite{Landau-QM-1977,Supplement}.

To model a bearing of the rotor shaft, we shall introduce the confinement potential,
\begin{equation}
  H_{\mathrm{pot}} = \frac{V}{2} \left( 1 - \cos \theta \right) ,
  \label{eq:HinderingPotential}
\end{equation}
where $V > 0$ is a system-dependent value; we may choose $V$ to be much larger than the typical energy scale of the rotation, as discussed later.
The potential term has a minimum value at $\theta = 0$, where $\theta$ is the angle between the $\zeta$- and $z$- axes.
We expect that this simple modeling for the bearing mechanism works well, at least, for a rough estiamte.
In the presence of the $H_\mathrm{pot}$, $L$ is no longer a conserved quantity~\cite{Supplement}.
In the following, we use $I_\xi = I_\eta = 8.71 \times 10^4 \ {\rm meV}~{\rm ps}^2$ and $I_\zeta = 4.29 \times 10^3 \ {\rm  meV}~{\rm ps}^2$, which are the values for an open-ended carbon nanotube defined by the two integers~\cite{RevModPhys.87.703} $(n,m)=(6,3)$ where the length $L_\mathrm{NT} = 4.75$ nm, and the potential $V = 1000$ meV.

The eigenstates and eigenenergies of the rotor under the potential can be calculated by diagonalizing the Hamiltonian matrices for $H_{R} + H_\mathrm{pot}$ constructed from the basis states $|L, M, k \rangle$ by utilizing the relations of the angular-momentum operators and the states~\cite{Supplement,Biedenharn1984} in each $(M,k)$ subspace.
To obtain the energy spectrum, it is convenient to consider an approximation taking the leading-order terms around $\theta = 0$ in the Hamiltonian.
The problem can then be mapped to that of a two-dimensional harmonic oscillator with the angular frequency $\omega = (V/2I_\xi)^{1/2}$~\cite{Supplement}.
Therefore, the analytical expression of the energy is given by
\begin{align}
  \varepsilon_{M, k, n} 
  = \left( 2 n + \left| M - k \right| + 1 \right) \hbar \omega
  + \frac{\hbar^2 k^2}{2 I_\zeta} 
  \label{eq:epsilon_Mkn}
\end{align}
where $n = 0,1,2,\ldots$ expresses the excitation in the harmonic oscillator for given $M$ and $k$.
The eigenenergies in Eq.~\eqref{eq:epsilon_Mkn} are shown as solid curves in Fig.~\ref{fig:en-k}.
The lowest branch $(M-k,n) = (0,0)$ corresponds to the motion of sleeping top, stable rotation around the $\zeta$ axis with the $\zeta$ axis aligned to the $z$ axis.
The upper branches shown in Fig.~\ref{fig:en-k} correspond to the precession state $(M-k,n) = (\pm 1,0)$ in which the angle between the $z$- and $\zeta$- axes is kept constant~\footnote{The higher branches that are not shown in Fig.~\ref{fig:en-k} include the nutation state in which $\theta$ fluctuates around an average angel $\theta_0 = \sqrt{ \hbar |M-k| / I_\xi \omega }$.
See Supplental Material for a detail.}.
When $\omega$ ($\propto \sqrt{V}$) is sufficiently large, the each branch is energetically well separated.

In the setup of electron transport, the electron wave function expands in the carbon nanotube then the nanorotor functions as a quantum dot
\footnote{
The recent quantum transport experiments with carefully fabricated carbon nanotube hybrid structures reveal the function of quantum dots~\cite{RevModPhys.87.703}.
In general, quantum levels in a nanotube are highly sensitively affected by contacted electrodes~\cite{Izumida-2015-06,Izumida-2016-05}.
However, we expect our rotor is an ideal carbon nanotube quantum dot since our proposed setup of the experiment is free from the contacts of the electrodes.}.
The number of electrons in the nanorotor changes one by one.
We shall model the electronic states in the nanorotor that contribute to transport with no ($n_d=0$) and a single electron ($n_d=1$).
These states can be represented with the spin index, i.e., $| s \rangle$, where $s=0$ for $n_d=0$ and $s = \pm 1/2$ for $n_d = 1$ [see Fig.~\ref{fig:geometry}(b)].
The Hamiltonian of the electronic states is then written as
$H_s = \sum_s \varepsilon_0 n_d(s) | s \rangle \langle s |$.
We shall choose $\varepsilon_0 = 0$ to represent the resonant tunneling condition.

In the nanorotor, an electron spin interacts with the mechanical rotational motion via the spin-rotation coupling, Eq.~\eqref{eq:spin-rotor-int}.
The quantum states of the nanorotor with a single electron are exactly obtained by diagonalizing the Hamiltonian matrices for $H_{R} + H_{\mathrm{pot}}+ H_s + H_\mathrm{SR}$.
The eigenenergies are plotted in Fig.~\ref{fig:en-k}, which shows a small splitting of the energies between the spin up and down states as a result of the interaction.
Note that the Hamiltonian conserves both $M+s$ and $k$.
Therefore, the spin-rotation coupling can mix the states in $(M, s)$ and $(M+2s, -s)$ subspaces.
That is, the interaction exchanges the $z$ component of the angular momentum between the electron and the mechanical rotation, but not the $\zeta$ component in the present setup for the symmetric top.
The angular-momentum exchange processes are diagrammatically shown in Fig.~\ref{fig:en-k} with the arrow between ($A$) and ($B$).
Note that the spin-rotation coupling can induce only the precession states.
To drive sleeping top motion efficiently, we will consider a relaxation process from the higher to the lowest branches [($B$) to ($C$) in Fig.~\ref{fig:en-k}] in the later discussion.

{\it Rotational motion induced by spin injection.---}
The effect of the electrodes on the rotor states is accounted for by the following Hamiltonian
\begin{align}
  H_e
  = \sum_{\substack{r=R,L\\q,s}} \varepsilon_{r q s} c_{r q s}^\dagger c_{r q s}
  + \sum_{\substack{r=R,L\\q,s}} V_{r q s} c_{r q s}^\dagger d_{s} + \mathrm{H.c.}
  \label{eq:lead}
\end{align}
The time evolution of the probability of the rotor states, $w_i$, is described by the master equation,
\begin{equation}
  \frac{d w_i}{dt} = \sum_{j} \Gamma_{ij} w_j - \sum_{j} \Gamma_{j i} w_i,
  \label{eq:MasterEq}
\end{equation}
where $i$ specifies an eigenstate of the rotor system including electronic states described by $H_{R} + H_{\mathrm{pot}} + H_s + H_\mathrm{SR}$~\cite{Supplement}.

As is often observed in macroscopic systems, relaxation processes would exist that lower the energy of the rotational motion and stabilize the motion.
To account for such processes, we explicitly consider a coupling to a bosonic bath modeled by the following Hamiltonian, 
\begin{align}
H_B 
= \sum_{q} \hbar \omega_{q} b_{q}^\dagger b_{q}
+ \sum_{q} g_{q} \left( b_{-q}^\dagger + b_{q} \right) L_+^{(\zeta)} + \mathrm{H.c.},
\label{eq:bath}
\end{align}
where $L_{+}^{(\zeta)}$ ($L_{-}^{(\zeta)}$) is the angular-momentum operator ascending (descending) the $k$ states,
\begin{align}
L_{\pm}^{(\zeta)} | L, M, k \rangle = \sqrt{ L(L+1) - k(k \pm 1)} |
L, M, k \pm 1 \rangle,
\end{align}
and the relaxation ratio is assumed to be
$\gamma^{(b)}(\varepsilon) = 2 \pi g \hbar^{-1} \varepsilon
\Theta(\varepsilon)$ for an Ohmic bath.  
As will be shown later, the process modeled by Eq.~\eqref{eq:bath} induces a relaxation from the precession state to the sleeping top state.
The effect of the relaxation expressed by Eq. \eqref{eq:bath} was clarified in a numerical simulation.

{\it Numerical simulation.---}
The time evolution of the rotational motion is calculated by solving the differential equation in Eq.~\eqref{eq:MasterEq}.
The temperature is set to $T= 10$ mK and the thermal equilibrium state of the rotor is chosen as the initial state.
The tunnel couplings are chosen as $\Gamma_{L \uparrow} = \Gamma_{R  \downarrow} = 5 \times 10^{-5}$ meV, $\Gamma_{L \downarrow} = \Gamma_{R \uparrow} = 0$ to represent the ferromagnetic electrodes.
The bias voltages are switched on at $t=0$ and the expected values $\langle M \rangle$ and $\langle k \rangle$ are calculated as a function of time.

\begin{figure}[tb]
  \includegraphics[width=8.5cm]{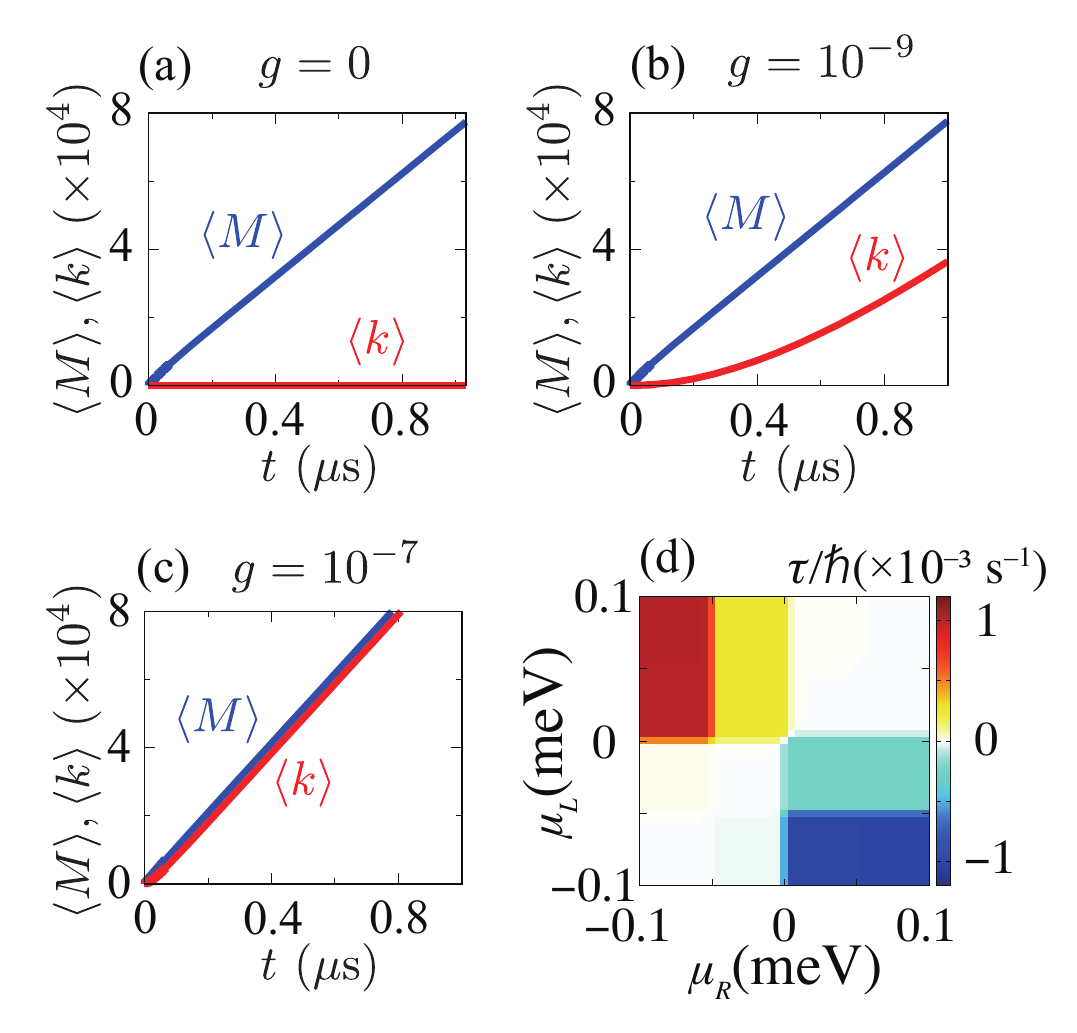}
  \caption{
    Time evolutions of rotational motion for (a) $g = 0$, (b) $g = 10^{-9}$, and (c) $g = 10^{-7}$.
    The chemical potentials of the left and right electrodes are chosen to be $\mu_L = -\mu_R = 0.1$ meV.
    (d) Density plot of torque in the source and drain chemical potentials, $\mu_L$-$\mu_R$, plane.}
  \label{fig:Lt}
\end{figure}

The time evolutions of the rotational motion in the absence of a relaxation ($g=0$) and in the presence of a relaxation ($g=10^{-9}$, $g=10^{-7}$) are shown in Figs.~\ref{fig:Lt}(a)-\ref{fig:Lt}(c).
In the absence of the relaxation, $\langle M \rangle$ increases almost linearly with respect to time, keeping $\langle k \rangle$ almost zero.
This means that the precession motion schematically represented in the upper inset of Fig.~\ref{fig:en-k} develops.
In contrast, in the presence of the relaxation, $\langle k \rangle$ increases with $\langle M \rangle$.
That is, the rotation in the sleeping top state $M=k$ that is schematically represented in the lower inset of Fig.~\ref{fig:en-k}(b) develops.
Note that the relaxation process indeed lowers the total energy by making a transition to the lowest branch.
These results clearly show the role of the relaxation in stabilizing the rotation to the sleeping top state.

As shown in Fig.~\ref{fig:Lt}(c), $\langle k \rangle$ develops linearly over a long time scale when the relaxation is present.
We shall consider the slope of $\langle k \rangle$ in time as follows;
\begin{equation}
\tau \equiv \hbar \frac{\langle k \rangle (t_\infty)}{t_\infty},
\end{equation}
where $t_\infty$ is taken to be a sufficiently long time after applying the bias voltage through which the time evolution develops linearly in time.
Note that $\tau$ is the torque to induce the sleeping top rotation.

\begin{figure}[tb]
  \includegraphics[width=8.5cm]{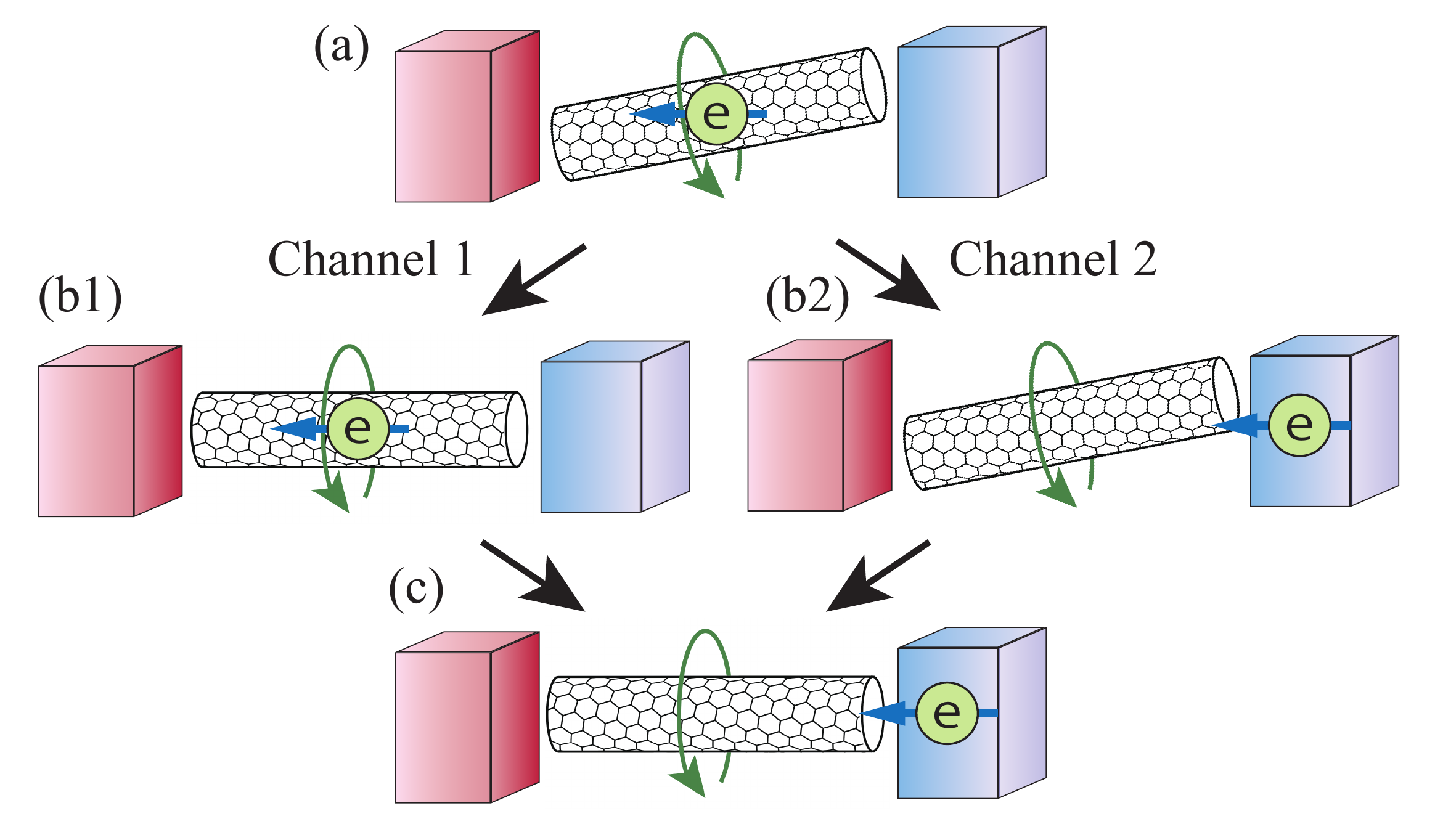}
  \caption{
    Two channels that are relevant to electron transfer.
}
  \label{fig:process}
\end{figure}

Figure \ref{fig:Lt}~(d) shows the calculated 
$\tau$
for wide range of source and drain bias voltages for $g = 10^{-7}$ and $t_\infty = 10 ~\mu$s.
Since the symmetry relation
$\tau(\mu_L,\mu_R)=-\tau(\mu_R,\mu_L)$
holds, let us focus on the positive bias regime $(\mu_L > \mu_R)$.
It is clear that there is a contribution to $\tau$ in the small bias regime, $-0.05 \ {\rm meV} < \mu_R < 0$, and it becomes large around the border $\mu_R \simeq -0.05 \ {\rm meV}$.
These behaviors can be understood as reflecting the number of current channels.
An electron with $s=1/2$ first tunnels from the left lead to the rotor, and the angler momentum transfer occurs because of the spin-rotation coupling: $(M, k, 1/2) \rightarrow (M+1, k, -1/2)$.
The resultant state is shown in Fig.~\ref{fig:process}(a).
In the small bias regime, the following process mainly contributes to the current flow, as shown by channel 1 in Fig.~\ref{fig:process}:
(i) Relaxation $(M+1, k, -1/2) \rightarrow (M+1, k+1, -1/2)$ and (ii) the spin-flipped electron (with $s=-1/2$) tunnels from the rotor to the right lead.
In addition to channel 1, channel 2 shown in Fig.~\ref{fig:process} also opens when the bias of the right electrode $\mu_R$ is lower than $- \hbar \omega$, which corresponds to the energy difference between the sleeping top and precession states.

Finally, we roughly estimate the angular velocity of the rotor by the condition that the torque of the spin-flip current balances with that of the friction in the double-wall nanotube~\cite{Supplement}.
By using the friction proportional to the relative velocity between the two tubes obtained by numerical simulations for double-walled carbon nanotubes~\cite{Servantie2006}, the steady-state angular velocity is estimated as $|\mathbf{\Omega}| \sim 100$ Hz.
The rotation would be directly confirmed by real-space probes such as the electron microscopes~\cite{Fennimore-Nature-2003,Sun-2005}.

{\it Summary.---}
We have proposed a nanoscale rotor embedded between two ferromagnetic electrodes that is driven by injected spins via the EdH effect.
As a feasible example, we considered a double-wall carbon nanotube rotor and developed a microscopic theory, in which the nanorotor
is modeled as an axisymmetric rigid body.
By solving the master equation, we found that the relaxation process from a precession state into a sleeping top state is crucial for efficient driving of the nanorotor.

The mechanism discussed in the Letter is not restricted to carbon nanotubes; it applies to other materials as well.
Further, there could exist other angular-momentum transfer processes from that of electron spin, such as spin-orbit and electron-phonon interactions, as well as nuclear spin interactions.
The locally deformed structure would eventually relax into rotational motion.
The process discussed in the Letter would be a minimal model for the momentum transfer from an electron to rotor that effectively includes the other possible processes.

\begin{acknowledgments}
 We acknowledge JSPS KAKENHI for Grants (No. JP16H01046, No. JP18H04282, No. JP19K14637, No. JP20K03831, No. JP20H01863, No. JP20K05258, and No. JP21K03414).
  M. M. is partially supported by the Priority Program of the Chinese Academy of Sciences, Grant No. XDB28000000.
\end{acknowledgments}


\begin{thebibliography}{41}%
\makeatletter
\providecommand \@ifxundefined [1]{%
 \@ifx{#1\undefined}
}%
\providecommand \@ifnum [1]{%
 \ifnum #1\expandafter \@firstoftwo
 \else \expandafter \@secondoftwo
 \fi
}%
\providecommand \@ifx [1]{%
 \ifx #1\expandafter \@firstoftwo
 \else \expandafter \@secondoftwo
 \fi
}%
\providecommand \natexlab [1]{#1}%
\providecommand \enquote  [1]{``#1''}%
\providecommand \bibnamefont  [1]{#1}%
\providecommand \bibfnamefont [1]{#1}%
\providecommand \citenamefont [1]{#1}%
\providecommand \href@noop [0]{\@secondoftwo}%
\providecommand \href [0]{\begingroup \@sanitize@url \@href}%
\providecommand \@href[1]{\@@startlink{#1}\@@href}%
\providecommand \@@href[1]{\endgroup#1\@@endlink}%
\providecommand \@sanitize@url [0]{\catcode `\\12\catcode `\$12\catcode
  `\&12\catcode `\#12\catcode `\^12\catcode `\_12\catcode `\%12\relax}%
\providecommand \@@startlink[1]{}%
\providecommand \@@endlink[0]{}%
\providecommand \url  [0]{\begingroup\@sanitize@url \@url }%
\providecommand \@url [1]{\endgroup\@href {#1}{\urlprefix }}%
\providecommand \urlprefix  [0]{URL }%
\providecommand \Eprint [0]{\href }%
\providecommand \doibase [0]{http://dx.doi.org/}%
\providecommand \selectlanguage [0]{\@gobble}%
\providecommand \bibinfo  [0]{\@secondoftwo}%
\providecommand \bibfield  [0]{\@secondoftwo}%
\providecommand \translation [1]{[#1]}%
\providecommand \BibitemOpen [0]{}%
\providecommand \bibitemStop [0]{}%
\providecommand \bibitemNoStop [0]{.\EOS\space}%
\providecommand \EOS [0]{\spacefactor3000\relax}%
\providecommand \BibitemShut  [1]{\csname bibitem#1\endcsname}%
\let\auto@bib@innerbib\@empty
\bibitem [{\citenamefont {Barnett}(1915)}]{Barnett1915}%
  \BibitemOpen
  \bibfield  {author} {\bibinfo {author} {\bibfnamefont {S.~J.}\ \bibnamefont
  {Barnett}},\ }\bibfield  {title} {\enquote {\bibinfo {title} {Magnetization
  by rotation},}\ }\href {\doibase 10.1103/PhysRev.6.239} {\bibfield  {journal}
  {\bibinfo  {journal} {Phys. Rev.}\ }\textbf {\bibinfo {volume} {6}},\
  \bibinfo {pages} {239} (\bibinfo {year} {1915})}\BibitemShut {NoStop}%
\bibitem [{\citenamefont {Einstein}\ and\ \citenamefont
  {de~Haas}(1915)}]{Einstein_deHaas_1915}%
  \BibitemOpen
  \bibfield  {author} {\bibinfo {author} {\bibfnamefont {A.}~\bibnamefont
  {Einstein}}\ and\ \bibinfo {author} {\bibfnamefont {W.~J.}\ \bibnamefont
  {de~Haas}},\ }\bibfield  {title} {\enquote {\bibinfo {title} {Experimental
  proof of the existence of amp\`{e}re's molecular currents},}\ }\href@noop {}
  {\bibfield  {journal} {\bibinfo  {journal} {KNAW proc.}\ }\textbf {\bibinfo
  {volume} {18 I}},\ \bibinfo {pages} {696} (\bibinfo {year}
  {1915})}\BibitemShut {NoStop}%
\bibitem [{\citenamefont {Scott}(1962)}]{Scott1962}%
  \BibitemOpen
  \bibfield  {author} {\bibinfo {author} {\bibfnamefont {G.~G.}\ \bibnamefont
  {Scott}},\ }\bibfield  {title} {\enquote {\bibinfo {title} {Review of
  gyromagnetic ratio experiments},}\ }\href {\doibase
  10.1103/RevModPhys.34.102} {\bibfield  {journal} {\bibinfo  {journal} {Rev.
  Mod. Phys.}\ }\textbf {\bibinfo {volume} {34}},\ \bibinfo {pages} {102}
  (\bibinfo {year} {1962})}\BibitemShut {NoStop}%
\bibitem [{\citenamefont {Kawaguchi}\ \emph {et~al.}(2006)\citenamefont
  {Kawaguchi}, \citenamefont {Saito},\ and\ \citenamefont
  {Ueda}}]{Kawaguchi2006}%
  \BibitemOpen
  \bibfield  {author} {\bibinfo {author} {\bibfnamefont {Y.}~\bibnamefont
  {Kawaguchi}}, \bibinfo {author} {\bibfnamefont {H.}~\bibnamefont {Saito}}, \
  and\ \bibinfo {author} {\bibfnamefont {M.}~\bibnamefont {Ueda}},\ }\bibfield
  {title} {\enquote {\bibinfo {title} {{E}instein--de {H}aas effect in dipolar
  {B}ose-{E}instein condensates},}\ }\href {\doibase
  10.1103/PhysRevLett.96.080405} {\bibfield  {journal} {\bibinfo  {journal}
  {Phys. Rev. Lett.}\ }\textbf {\bibinfo {volume} {96}},\ \bibinfo {pages}
  {080405} (\bibinfo {year} {2006})}\BibitemShut {NoStop}%
\bibitem [{\citenamefont {Gawryluk}\ \emph {et~al.}(2007)\citenamefont
  {Gawryluk}, \citenamefont {Brewczyk}, \citenamefont {Bongs},\ and\
  \citenamefont {Gajda}}]{Gawryluk2007}%
  \BibitemOpen
  \bibfield  {author} {\bibinfo {author} {\bibfnamefont {K.}~\bibnamefont
  {Gawryluk}}, \bibinfo {author} {\bibfnamefont {M.}~\bibnamefont {Brewczyk}},
  \bibinfo {author} {\bibfnamefont {K.}~\bibnamefont {Bongs}}, \ and\ \bibinfo
  {author} {\bibfnamefont {M.}~\bibnamefont {Gajda}},\ }\bibfield  {title}
  {\enquote {\bibinfo {title} {Resonant {E}instein--de {H}aas effect in a
  rubidium condensate},}\ }\href {\doibase 10.1103/PhysRevLett.99.130401}
  {\bibfield  {journal} {\bibinfo  {journal} {Phys. Rev. Lett.}\ }\textbf
  {\bibinfo {volume} {99}},\ \bibinfo {pages} {130401} (\bibinfo {year}
  {2007})}\BibitemShut {NoStop}%
\bibitem [{\citenamefont {Takahashi}\ \emph {et~al.}(2016)\citenamefont
  {Takahashi}, \citenamefont {Matsuo}, \citenamefont {Ono}, \citenamefont
  {Harii}, \citenamefont {Chudo}, \citenamefont {Okayasu}, \citenamefont
  {Ieda}, \citenamefont {Takahashi}, \citenamefont {Maekawa},\ and\
  \citenamefont {Saitoh}}]{Takahashi2016}%
  \BibitemOpen
  \bibfield  {author} {\bibinfo {author} {\bibfnamefont {R.}~\bibnamefont
  {Takahashi}}, \bibinfo {author} {\bibfnamefont {M.}~\bibnamefont {Matsuo}},
  \bibinfo {author} {\bibfnamefont {M.}~\bibnamefont {Ono}}, \bibinfo {author}
  {\bibfnamefont {K.}~\bibnamefont {Harii}}, \bibinfo {author} {\bibfnamefont
  {H.}~\bibnamefont {Chudo}}, \bibinfo {author} {\bibfnamefont
  {S.}~\bibnamefont {Okayasu}}, \bibinfo {author} {\bibfnamefont
  {J.}~\bibnamefont {Ieda}}, \bibinfo {author} {\bibfnamefont {S.}~\bibnamefont
  {Takahashi}}, \bibinfo {author} {\bibfnamefont {S.}~\bibnamefont {Maekawa}},
  \ and\ \bibinfo {author} {\bibfnamefont {E.}~\bibnamefont {Saitoh}},\
  }\bibfield  {title} {\enquote {\bibinfo {title} {Spin hydrodynamic
  generation},}\ }\href {\doibase 10.1038/nphys3526} {\bibfield  {journal}
  {\bibinfo  {journal} {Nat. Phys.}\ }\textbf {\bibinfo {volume} {12}},\
  \bibinfo {pages} {52} (\bibinfo {year} {2016})}\BibitemShut {NoStop}%
\bibitem [{\citenamefont {Kobayashi}\ \emph {et~al.}(2017)\citenamefont
  {Kobayashi}, \citenamefont {Yoshikawa}, \citenamefont {Matsuo}, \citenamefont
  {Iguchi}, \citenamefont {Maekawa}, \citenamefont {Saitoh},\ and\
  \citenamefont {Nozaki}}]{Kobayashi2017}%
  \BibitemOpen
  \bibfield  {author} {\bibinfo {author} {\bibfnamefont {D.}~\bibnamefont
  {Kobayashi}}, \bibinfo {author} {\bibfnamefont {T.}~\bibnamefont
  {Yoshikawa}}, \bibinfo {author} {\bibfnamefont {M.}~\bibnamefont {Matsuo}},
  \bibinfo {author} {\bibfnamefont {R.}~\bibnamefont {Iguchi}}, \bibinfo
  {author} {\bibfnamefont {S.}~\bibnamefont {Maekawa}}, \bibinfo {author}
  {\bibfnamefont {E.}~\bibnamefont {Saitoh}}, \ and\ \bibinfo {author}
  {\bibfnamefont {Y.}~\bibnamefont {Nozaki}},\ }\bibfield  {title} {\enquote
  {\bibinfo {title} {Spin current generation using a surface acoustic wave
  generated via spin-rotation coupling},}\ }\href {\doibase
  10.1103/PhysRevLett.119.077202} {\bibfield  {journal} {\bibinfo  {journal}
  {Phys. Rev. Lett.}\ }\textbf {\bibinfo {volume} {119}},\ \bibinfo {pages}
  {077202} (\bibinfo {year} {2017})}\BibitemShut {NoStop}%
\bibitem [{\citenamefont {Kurimune}\ \emph {et~al.}(2020)\citenamefont
  {Kurimune}, \citenamefont {Matsuo},\ and\ \citenamefont
  {Nozaki}}]{Kurimune2020}%
  \BibitemOpen
  \bibfield  {author} {\bibinfo {author} {\bibfnamefont {Y.}~\bibnamefont
  {Kurimune}}, \bibinfo {author} {\bibfnamefont {M.}~\bibnamefont {Matsuo}}, \
  and\ \bibinfo {author} {\bibfnamefont {Y.}~\bibnamefont {Nozaki}},\
  }\bibfield  {title} {\enquote {\bibinfo {title} {Observation of gyromagnetic
  spin wave resonance in nife films},}\ }\href {\doibase
  10.1103/PhysRevLett.124.217205} {\bibfield  {journal} {\bibinfo  {journal}
  {Phys. Rev. Lett.}\ }\textbf {\bibinfo {volume} {124}},\ \bibinfo {pages}
  {217205} (\bibinfo {year} {2020})}\BibitemShut {NoStop}%
\bibitem [{\citenamefont {Takahashi}\ \emph {et~al.}(2020)\citenamefont
  {Takahashi}, \citenamefont {Chudo}, \citenamefont {Matsuo}, \citenamefont
  {Harii}, \citenamefont {Ohnuma}, \citenamefont {Maekawa},\ and\ \citenamefont
  {Saitoh}}]{Takahashi2020}%
  \BibitemOpen
  \bibfield  {author} {\bibinfo {author} {\bibfnamefont {R.}~\bibnamefont
  {Takahashi}}, \bibinfo {author} {\bibfnamefont {H.}~\bibnamefont {Chudo}},
  \bibinfo {author} {\bibfnamefont {M.}~\bibnamefont {Matsuo}}, \bibinfo
  {author} {\bibfnamefont {K.}~\bibnamefont {Harii}}, \bibinfo {author}
  {\bibfnamefont {Y.}~\bibnamefont {Ohnuma}}, \bibinfo {author} {\bibfnamefont
  {S.}~\bibnamefont {Maekawa}}, \ and\ \bibinfo {author} {\bibfnamefont
  {E.}~\bibnamefont {Saitoh}},\ }\bibfield  {title} {\enquote {\bibinfo {title}
  {Giant spin hydrodynamic generation in laminar flow},}\ }\href {\doibase
  10.1038/s41467-020-16753-0} {\bibfield  {journal} {\bibinfo  {journal} {Nat.
  Commun.}\ }\textbf {\bibinfo {volume} {11}},\ \bibinfo {pages} {3009}
  (\bibinfo {year} {2020})}\BibitemShut {NoStop}%
\bibitem [{\citenamefont {Kazerooni}\ \emph {et~al.}(2020)\citenamefont
  {Kazerooni}, \citenamefont {Thieme}, \citenamefont {Schumacher},\ and\
  \citenamefont {Cierpka}}]{Kazerooni2020}%
  \BibitemOpen
  \bibfield  {author} {\bibinfo {author} {\bibfnamefont {T.~H.}\ \bibnamefont
  {Kazerooni}}, \bibinfo {author} {\bibfnamefont {A.}~\bibnamefont {Thieme}},
  \bibinfo {author} {\bibfnamefont {J.}~\bibnamefont {Schumacher}}, \ and\
  \bibinfo {author} {\bibfnamefont {C.}~\bibnamefont {Cierpka}},\ }\bibfield
  {title} {\enquote {\bibinfo {title} {Electron spin-vorticity coupling in pipe
  flows at low and high reynolds number},}\ }\href {\doibase
  10.1103/PhysRevApplied.14.014002} {\bibfield  {journal} {\bibinfo  {journal}
  {Phys. Rev. Applied}\ }\textbf {\bibinfo {volume} {14}},\ \bibinfo {pages}
  {014002} (\bibinfo {year} {2020})}\BibitemShut {NoStop}%
\bibitem [{\citenamefont {Kazerooni}\ \emph {et~al.}(2021)\citenamefont
  {Kazerooni}, \citenamefont {Zinchenko}, \citenamefont {Schumacher},\ and\
  \citenamefont {Cierpka}}]{Kazerooni2021}%
  \BibitemOpen
  \bibfield  {author} {\bibinfo {author} {\bibfnamefont {T.~H.}\ \bibnamefont
  {Kazerooni}}, \bibinfo {author} {\bibfnamefont {G.}~\bibnamefont
  {Zinchenko}}, \bibinfo {author} {\bibfnamefont {J.}~\bibnamefont
  {Schumacher}}, \ and\ \bibinfo {author} {\bibfnamefont {C.}~\bibnamefont
  {Cierpka}},\ }\bibfield  {title} {\enquote {\bibinfo {title} {Electrical
  voltage by electron spin-vorticity coupling in laminar ducts},}\ }\href
  {\doibase 10.1103/PhysRevFluids.6.043703} {\bibfield  {journal} {\bibinfo
  {journal} {Phys. Rev. Fluids}\ }\textbf {\bibinfo {volume} {6}},\ \bibinfo
  {pages} {043703} (\bibinfo {year} {2021})}\BibitemShut {NoStop}%
\bibitem [{\citenamefont {Chudo}\ \emph {et~al.}(2014)\citenamefont {Chudo},
  \citenamefont {Ono}, \citenamefont {Harii}, \citenamefont {Matsuo},
  \citenamefont {Ieda}, \citenamefont {Haruki}, \citenamefont {Okayasu},
  \citenamefont {Maekawa}, \citenamefont {Yasuoka},\ and\ \citenamefont
  {Saitoh}}]{Chudo2014}%
  \BibitemOpen
  \bibfield  {author} {\bibinfo {author} {\bibfnamefont {H.}~\bibnamefont
  {Chudo}}, \bibinfo {author} {\bibfnamefont {M.}~\bibnamefont {Ono}}, \bibinfo
  {author} {\bibfnamefont {K.}~\bibnamefont {Harii}}, \bibinfo {author}
  {\bibfnamefont {M.}~\bibnamefont {Matsuo}}, \bibinfo {author} {\bibfnamefont
  {J.}~\bibnamefont {Ieda}}, \bibinfo {author} {\bibfnamefont {R.}~\bibnamefont
  {Haruki}}, \bibinfo {author} {\bibfnamefont {S.}~\bibnamefont {Okayasu}},
  \bibinfo {author} {\bibfnamefont {S.}~\bibnamefont {Maekawa}}, \bibinfo
  {author} {\bibfnamefont {H.}~\bibnamefont {Yasuoka}}, \ and\ \bibinfo
  {author} {\bibfnamefont {E.}~\bibnamefont {Saitoh}},\ }\bibfield  {title}
  {\enquote {\bibinfo {title} {Observation of {B}arnett fields in solids by
  nuclear magnetic resonance},}\ }\href {\doibase 10.7567/apex.7.063004}
  {\bibfield  {journal} {\bibinfo  {journal} {Appl. Phys. Express}\ }\textbf
  {\bibinfo {volume} {7}},\ \bibinfo {pages} {063004} (\bibinfo {year}
  {2014})}\BibitemShut {NoStop}%
\bibitem [{\citenamefont {Collaboration}(2017)}]{Adamczyk2017}%
  \BibitemOpen
  \bibfield  {author} {\bibinfo {author} {\bibfnamefont {The~STAR}\
  \bibnamefont {Collaboration}},\ }\bibfield  {title} {\enquote {\bibinfo
  {title} {{Global $\Lambda$ hyperon polarization in nuclear collisions}},}\
  }\href {\doibase 10.1038/nature23004} {\bibfield  {journal} {\bibinfo
  {journal} {Nature}\ }\textbf {\bibinfo {volume} {548}},\ \bibinfo {pages}
  {62} (\bibinfo {year} {2017})}\BibitemShut {NoStop}%
\bibitem [{\citenamefont {Dornes}\ \emph {et~al.}(2019)\citenamefont {Dornes},
  \citenamefont {Acremann}, \citenamefont {Savoini}, \citenamefont {Kubli},
  \citenamefont {Neugebauer}, \citenamefont {Abreu}, \citenamefont {Huber},
  \citenamefont {Lantz}, \citenamefont {Vaz}, \citenamefont {Lemke},
  \citenamefont {Bothschafter}, \citenamefont {Porer}, \citenamefont
  {Esposito}, \citenamefont {Rettig}, \citenamefont {Buzzi}, \citenamefont
  {Alberca}, \citenamefont {Windsor}, \citenamefont {Beaud}, \citenamefont
  {Staub}, \citenamefont {Zhu}, \citenamefont {Song}, \citenamefont {Glownia},\
  and\ \citenamefont {Johnson}}]{Dornes2019}%
  \BibitemOpen
  \bibfield  {author} {\bibinfo {author} {\bibfnamefont {C.}~\bibnamefont
  {Dornes}}, \bibinfo {author} {\bibfnamefont {Y.}~\bibnamefont {Acremann}},
  \bibinfo {author} {\bibfnamefont {M.}~\bibnamefont {Savoini}}, \bibinfo
  {author} {\bibfnamefont {M.}~\bibnamefont {Kubli}}, \bibinfo {author}
  {\bibfnamefont {M.~J.}\ \bibnamefont {Neugebauer}}, \bibinfo {author}
  {\bibfnamefont {E.}~\bibnamefont {Abreu}}, \bibinfo {author} {\bibfnamefont
  {L.}~\bibnamefont {Huber}}, \bibinfo {author} {\bibfnamefont
  {G.}~\bibnamefont {Lantz}}, \bibinfo {author} {\bibfnamefont {C.~A.~F.}\
  \bibnamefont {Vaz}}, \bibinfo {author} {\bibfnamefont {H.}~\bibnamefont
  {Lemke}}, \bibinfo {author} {\bibfnamefont {E.~M.}\ \bibnamefont
  {Bothschafter}}, \bibinfo {author} {\bibfnamefont {M}~\bibnamefont {Porer}},
  \bibinfo {author} {\bibfnamefont {V.}~\bibnamefont {Esposito}}, \bibinfo
  {author} {\bibfnamefont {L.}~\bibnamefont {Rettig}}, \bibinfo {author}
  {\bibfnamefont {M.}~\bibnamefont {Buzzi}}, \bibinfo {author} {\bibfnamefont
  {A.}~\bibnamefont {Alberca}}, \bibinfo {author} {\bibfnamefont {Y.~W.}\
  \bibnamefont {Windsor}}, \bibinfo {author} {\bibfnamefont {P.}~\bibnamefont
  {Beaud}}, \bibinfo {author} {\bibfnamefont {U.}~\bibnamefont {Staub}},
  \bibinfo {author} {\bibfnamefont {Diling}\ \bibnamefont {Zhu}}, \bibinfo
  {author} {\bibfnamefont {Sanghoon}\ \bibnamefont {Song}}, \bibinfo {author}
  {\bibfnamefont {J.~M.}\ \bibnamefont {Glownia}}, \ and\ \bibinfo {author}
  {\bibfnamefont {S.~L.}\ \bibnamefont {Johnson}},\ }\bibfield  {title}
  {\enquote {\bibinfo {title} {{The ultrafast {E}instein^^e2^^80^^93de {H}aas
  effect}},}\ }\href {\doibase 10.1038/s41586-018-0822-7} {\bibfield  {journal}
  {\bibinfo  {journal} {Nature}\ }\textbf {\bibinfo {volume} {565}},\ \bibinfo
  {pages} {209--212} (\bibinfo {year} {2019})}\BibitemShut {NoStop}%
\bibitem [{\citenamefont {Mohanty}\ \emph {et~al.}(2004)\citenamefont
  {Mohanty}, \citenamefont {Zolfagharkhani}, \citenamefont {Kettemann},\ and\
  \citenamefont {Fulde}}]{Mohanty-PRB-2004}%
  \BibitemOpen
  \bibfield  {author} {\bibinfo {author} {\bibfnamefont {P.}~\bibnamefont
  {Mohanty}}, \bibinfo {author} {\bibfnamefont {G.}~\bibnamefont
  {Zolfagharkhani}}, \bibinfo {author} {\bibfnamefont {S.}~\bibnamefont
  {Kettemann}}, \ and\ \bibinfo {author} {\bibfnamefont {P.}~\bibnamefont
  {Fulde}},\ }\bibfield  {title} {\enquote {\bibinfo {title} {Spin-mechanical
  device for detection and control of spin current by nanomechanical torque},}\
  }\href {\doibase 10.1103/PhysRevB.70.195301} {\bibfield  {journal} {\bibinfo
  {journal} {Phys. Rev. B}\ }\textbf {\bibinfo {volume} {70}},\ \bibinfo
  {pages} {195301} (\bibinfo {year} {2004})}\BibitemShut {NoStop}%
\bibitem [{\citenamefont {Wallis}\ \emph {et~al.}(2006)\citenamefont {Wallis},
  \citenamefont {Moreland},\ and\ \citenamefont {Kabos}}]{Wallis-APL-2006-09}%
  \BibitemOpen
  \bibfield  {author} {\bibinfo {author} {\bibfnamefont {T.~M.}\ \bibnamefont
  {Wallis}}, \bibinfo {author} {\bibfnamefont {J.}~\bibnamefont {Moreland}}, \
  and\ \bibinfo {author} {\bibfnamefont {P.}~\bibnamefont {Kabos}},\ }\bibfield
   {title} {\enquote {\bibinfo {title} {{E}instein--de {H}aas effect in a
  {N}i{F}e film deposited on a microcantilever},}\ }\href@noop {} {\bibfield
  {journal} {\bibinfo  {journal} {Appl. Phys. Lett.}\ }\textbf {\bibinfo
  {volume} {89}},\ \bibinfo {pages} {122502} (\bibinfo {year}
  {2006})}\BibitemShut {NoStop}%
\bibitem [{\citenamefont {Zolfagharkhani}\ \emph {et~al.}(2008)\citenamefont
  {Zolfagharkhani}, \citenamefont {Gaidarzhy}, \citenamefont {Degiovanni},
  \citenamefont {Kettemann}, \citenamefont {Fulde},\ and\ \citenamefont
  {Mohanty}}]{Zolfagharkhani-NatNano-2008}%
  \BibitemOpen
  \bibfield  {author} {\bibinfo {author} {\bibfnamefont {G.}~\bibnamefont
  {Zolfagharkhani}}, \bibinfo {author} {\bibfnamefont {A.}~\bibnamefont
  {Gaidarzhy}}, \bibinfo {author} {\bibfnamefont {P.}~\bibnamefont
  {Degiovanni}}, \bibinfo {author} {\bibfnamefont {S.}~\bibnamefont
  {Kettemann}}, \bibinfo {author} {\bibfnamefont {P.}~\bibnamefont {Fulde}}, \
  and\ \bibinfo {author} {\bibfnamefont {P.}~\bibnamefont {Mohanty}},\
  }\bibfield  {title} {\enquote {\bibinfo {title} {Nanomechanical detection of
  itinerant electron spin flip},}\ }\href@noop {} {\bibfield  {journal}
  {\bibinfo  {journal} {Nat. Nanotechnol.}\ }\textbf {\bibinfo {volume} {3}},\
  \bibinfo {pages} {720--723} (\bibinfo {year} {2008})}\BibitemShut {NoStop}%
\bibitem [{\citenamefont {Harii}\ \emph {et~al.}(2019)\citenamefont {Harii},
  \citenamefont {Seo}, \citenamefont {Tsutsumi}, \citenamefont {Chudo},
  \citenamefont {Oyanagi}, \citenamefont {Matsuo}, \citenamefont {Shiomi},
  \citenamefont {Ono}, \citenamefont {Maekawa},\ and\ \citenamefont
  {Saitoh}}]{Harii-2019-NatCommun}%
  \BibitemOpen
  \bibfield  {author} {\bibinfo {author} {\bibfnamefont {K.}~\bibnamefont
  {Harii}}, \bibinfo {author} {\bibfnamefont {Y.-J.}\ \bibnamefont {Seo}},
  \bibinfo {author} {\bibfnamefont {Y.}~\bibnamefont {Tsutsumi}}, \bibinfo
  {author} {\bibfnamefont {H.}~\bibnamefont {Chudo}}, \bibinfo {author}
  {\bibfnamefont {K.}~\bibnamefont {Oyanagi}}, \bibinfo {author} {\bibfnamefont
  {M.}~\bibnamefont {Matsuo}}, \bibinfo {author} {\bibfnamefont
  {Y.}~\bibnamefont {Shiomi}}, \bibinfo {author} {\bibfnamefont
  {T.}~\bibnamefont {Ono}}, \bibinfo {author} {\bibfnamefont {S.}~\bibnamefont
  {Maekawa}}, \ and\ \bibinfo {author} {\bibfnamefont {E.}~\bibnamefont
  {Saitoh}},\ }\bibfield  {title} {\enquote {\bibinfo {title} {Spin {S}eebeck
  mechanical force},}\ }\href {\doibase 10.1038/s41467-019-10625-y} {\bibfield
  {journal} {\bibinfo  {journal} {Nat. Commun.}\ }\textbf {\bibinfo {volume}
  {10}},\ \bibinfo {pages} {2616} (\bibinfo {year} {2019})}\BibitemShut
  {NoStop}%
\bibitem [{\citenamefont {Cleland}(2003)}]{Cleland-2003-book}%
  \BibitemOpen
  \bibfield  {author} {\bibinfo {author} {\bibfnamefont {A.~N.}\ \bibnamefont
  {Cleland}},\ }\href@noop {} {\emph {\bibinfo {title} {Foundations of
  Nanomechanics: From Solid-State Theory to Device Applications}}}\ (\bibinfo
  {publisher} {Springer},\ \bibinfo {address} {Berlin, Germany},\ \bibinfo
  {year} {2003})\BibitemShut {NoStop}%
\bibitem [{\citenamefont {Fennimore}\ \emph {et~al.}(2003)\citenamefont
  {Fennimore}, \citenamefont {Yuzvinsky}, \citenamefont {Han}, \citenamefont
  {Fuhrer}, \citenamefont {Cumings},\ and\ \citenamefont
  {Zettl}}]{Fennimore-Nature-2003}%
  \BibitemOpen
  \bibfield  {author} {\bibinfo {author} {\bibfnamefont {A.~M.}\ \bibnamefont
  {Fennimore}}, \bibinfo {author} {\bibfnamefont {T.~D.}\ \bibnamefont
  {Yuzvinsky}}, \bibinfo {author} {\bibfnamefont {W.-Q.}\ \bibnamefont {Han}},
  \bibinfo {author} {\bibfnamefont {M.~S.}\ \bibnamefont {Fuhrer}}, \bibinfo
  {author} {\bibfnamefont {J.}~\bibnamefont {Cumings}}, \ and\ \bibinfo
  {author} {\bibfnamefont {A.}~\bibnamefont {Zettl}},\ }\bibfield  {title}
  {\enquote {\bibinfo {title} {Rotational actuators based on carbon
  nanotubes},}\ }\href@noop {} {\bibfield  {journal} {\bibinfo  {journal}
  {Nature}\ }\textbf {\bibinfo {volume} {424}},\ \bibinfo {pages} {408--410}
  (\bibinfo {year} {2003})}\BibitemShut {NoStop}%
\bibitem [{\citenamefont {Bailey}\ \emph {et~al.}(2008)\citenamefont {Bailey},
  \citenamefont {Amanatidis},\ and\ \citenamefont {Lambert}}]{Bailey-PRL-2008}%
  \BibitemOpen
  \bibfield  {author} {\bibinfo {author} {\bibfnamefont {S.~W.~D.}\
  \bibnamefont {Bailey}}, \bibinfo {author} {\bibfnamefont {I.}~\bibnamefont
  {Amanatidis}}, \ and\ \bibinfo {author} {\bibfnamefont {C.~J.}\ \bibnamefont
  {Lambert}},\ }\bibfield  {title} {\enquote {\bibinfo {title} {Carbon nanotube
  electron windmills: A novel design for nanomotors},}\ }\href {\doibase
  10.1103/PhysRevLett.100.256802} {\bibfield  {journal} {\bibinfo  {journal}
  {Phys. Rev. Lett.}\ }\textbf {\bibinfo {volume} {100}},\ \bibinfo {pages}
  {256802} (\bibinfo {year} {2008})}\BibitemShut {NoStop}%
\bibitem [{\citenamefont {Iijima}(1991)}]{Iijima-1991}%
  \BibitemOpen
  \bibfield  {author} {\bibinfo {author} {\bibfnamefont {S.}~\bibnamefont
  {Iijima}},\ }\bibfield  {title} {\enquote {\bibinfo {title} {Helical
  microtubules of graphitic carbon},}\ }\href {\doibase 10.1038/354056a0}
  {\bibfield  {journal} {\bibinfo  {journal} {Nature}\ }\textbf {\bibinfo
  {volume} {354}},\ \bibinfo {pages} {56--58} (\bibinfo {year}
  {1991})}\BibitemShut {NoStop}%
\bibitem [{\citenamefont {Smith}\ \emph {et~al.}(1998)\citenamefont {Smith},
  \citenamefont {Monthioux},\ and\ \citenamefont {Luzzi}}]{Smith-nature-1998}%
  \BibitemOpen
  \bibfield  {author} {\bibinfo {author} {\bibfnamefont {B.~W.}\ \bibnamefont
  {Smith}}, \bibinfo {author} {\bibfnamefont {M.}~\bibnamefont {Monthioux}}, \
  and\ \bibinfo {author} {\bibfnamefont {D.~E.}\ \bibnamefont {Luzzi}},\
  }\bibfield  {title} {\enquote {\bibinfo {title} {Encapsulated {C}60 in carbon
  nanotubes},}\ }\href {\doibase 10.1038/24521} {\bibfield  {journal} {\bibinfo
   {journal} {Nature}\ }\textbf {\bibinfo {volume} {396}},\ \bibinfo {pages}
  {323--324} (\bibinfo {year} {1998})}\BibitemShut {NoStop}%
\bibitem [{\citenamefont {Cumings}\ and\ \citenamefont
  {Zettl}(2000)}]{Cumings-Zettl-2000}%
  \BibitemOpen
  \bibfield  {author} {\bibinfo {author} {\bibfnamefont {J.}~\bibnamefont
  {Cumings}}\ and\ \bibinfo {author} {\bibfnamefont {A.}~\bibnamefont
  {Zettl}},\ }\bibfield  {title} {\enquote {\bibinfo {title} {Low-friction
  nanoscale linear bearing realized from multiwall carbon nanotubes},}\ }\href
  {\doibase 10.1126/science.289.5479.602} {\bibfield  {journal} {\bibinfo
  {journal} {Science}\ }\textbf {\bibinfo {volume} {289}},\ \bibinfo {pages}
  {602--604} (\bibinfo {year} {2000})}\BibitemShut {NoStop}%
\bibitem [{\citenamefont {Cook}\ \emph {et~al.}(2013)\citenamefont {Cook},
  \citenamefont {Buehler},\ and\ \citenamefont {Spakovszky}}]{Cook2013}%
  \BibitemOpen
  \bibfield  {author} {\bibinfo {author} {\bibfnamefont {E.~H.}\ \bibnamefont
  {Cook}}, \bibinfo {author} {\bibfnamefont {M.~J.}\ \bibnamefont {Buehler}}, \
  and\ \bibinfo {author} {\bibfnamefont {Z.~S.}\ \bibnamefont {Spakovszky}},\
  }\bibfield  {title} {\enquote {\bibinfo {title} {Mechanism of friction in
  rotating carbon nanotube bearings},}\ }\href {\doibase
  https://doi.org/10.1016/j.jmps.2012.08.004} {\bibfield  {journal} {\bibinfo
  {journal} {J. Mech. Phys. Solids}\ }\textbf {\bibinfo {volume} {61}},\
  \bibinfo {pages} {652} (\bibinfo {year} {2013})}\BibitemShut {NoStop}%
\bibitem [{\citenamefont {Huang}\ and\ \citenamefont {Han}(2018)}]{Huang2018}%
  \BibitemOpen
  \bibfield  {author} {\bibinfo {author} {\bibfnamefont {J.}~\bibnamefont
  {Huang}}\ and\ \bibinfo {author} {\bibfnamefont {Q.}~\bibnamefont {Han}},\
  }\bibfield  {title} {\enquote {\bibinfo {title} {Strain effects on rotational
  property in nanoscale rotation system},}\ }\href {\doibase
  10.1038/s41598-017-18903-9} {\bibfield  {journal} {\bibinfo  {journal} {Sci.
  Rep.}\ }\textbf {\bibinfo {volume} {8}},\ \bibinfo {pages} {432} (\bibinfo
  {year} {2018})}\BibitemShut {NoStop}%
\bibitem [{foo()}]{footnote1}%
  \BibitemOpen
  \href@noop {} {}\bibinfo {howpublished} {Although other materials may be
  considered as a candidate for the nanorotor such as molecules and nanorods,
  we do not discuss such a possibility in this Letter.}\BibitemShut {Stop}%
\bibitem [{\citenamefont {Cottet}\ \emph {et~al.}(2006)\citenamefont {Cottet},
  \citenamefont {Kontos}, \citenamefont {Sahoo}, \citenamefont {Man},
  \citenamefont {Choi}, \citenamefont {Belzig}, \citenamefont {Bruder},
  \citenamefont {Morpurgo},\ and\ \citenamefont
  {Sch{\"o}nenberger}}]{Cottet-2006-10}%
  \BibitemOpen
  \bibfield  {author} {\bibinfo {author} {\bibfnamefont {A.}~\bibnamefont
  {Cottet}}, \bibinfo {author} {\bibfnamefont {T.}~\bibnamefont {Kontos}},
  \bibinfo {author} {\bibfnamefont {S.}~\bibnamefont {Sahoo}}, \bibinfo
  {author} {\bibfnamefont {H.~T.}\ \bibnamefont {Man}}, \bibinfo {author}
  {\bibfnamefont {M.-S.}\ \bibnamefont {Choi}}, \bibinfo {author}
  {\bibfnamefont {W.}~\bibnamefont {Belzig}}, \bibinfo {author} {\bibfnamefont
  {C.}~\bibnamefont {Bruder}}, \bibinfo {author} {\bibfnamefont {A.~F.}\
  \bibnamefont {Morpurgo}}, \ and\ \bibinfo {author} {\bibfnamefont
  {C.}~\bibnamefont {Sch{\"o}nenberger}},\ }\bibfield  {title} {\enquote
  {\bibinfo {title} {Nanospintronics with carbon nanotubes},}\ }\href@noop {}
  {\bibfield  {journal} {\bibinfo  {journal} {Semicond. Sci. Technol.}\
  }\textbf {\bibinfo {volume} {21}},\ \bibinfo {pages} {S78 -- S95} (\bibinfo
  {year} {2006})}\BibitemShut {NoStop}%
\bibitem [{\citenamefont {Laird}\ \emph {et~al.}(2015)\citenamefont {Laird},
  \citenamefont {Kuemmeth}, \citenamefont {Steele}, \citenamefont
  {Grove-Rasmussen}, \citenamefont {Nyg\aa{}rd}, \citenamefont {Flensberg},\
  and\ \citenamefont {Kouwenhoven}}]{RevModPhys.87.703}%
  \BibitemOpen
  \bibfield  {author} {\bibinfo {author} {\bibfnamefont {E.~A.}\ \bibnamefont
  {Laird}}, \bibinfo {author} {\bibfnamefont {F.}~\bibnamefont {Kuemmeth}},
  \bibinfo {author} {\bibfnamefont {G.~A.}\ \bibnamefont {Steele}}, \bibinfo
  {author} {\bibfnamefont {K.}~\bibnamefont {Grove-Rasmussen}}, \bibinfo
  {author} {\bibfnamefont {J.}~\bibnamefont {Nyg\aa{}rd}}, \bibinfo {author}
  {\bibfnamefont {K.}~\bibnamefont {Flensberg}}, \ and\ \bibinfo {author}
  {\bibfnamefont {L.~P.}\ \bibnamefont {Kouwenhoven}},\ }\bibfield  {title}
  {\enquote {\bibinfo {title} {Quantum transport in carbon nanotubes},}\
  }\href@noop {} {\bibfield  {journal} {\bibinfo  {journal} {Rev. Mod. Phys.}\
  }\textbf {\bibinfo {volume} {87}},\ \bibinfo {pages} {703} (\bibinfo {year}
  {2015})}\BibitemShut {NoStop}%
\bibitem [{\citenamefont {Izumida}\ \emph {et~al.}(2015)\citenamefont
  {Izumida}, \citenamefont {Okuyama},\ and\ \citenamefont
  {Saito}}]{Izumida-2015-06}%
  \BibitemOpen
  \bibfield  {author} {\bibinfo {author} {\bibfnamefont {W.}~\bibnamefont
  {Izumida}}, \bibinfo {author} {\bibfnamefont {R.}~\bibnamefont {Okuyama}}, \
  and\ \bibinfo {author} {\bibfnamefont {R.}~\bibnamefont {Saito}},\ }\bibfield
   {title} {\enquote {\bibinfo {title} {Valley coupling in finite-length
  metallic single-wall carbon nanotubes},}\ }\href {\doibase
  10.1103/PhysRevB.91.235442} {\bibfield  {journal} {\bibinfo  {journal} {Phys.
  Rev. B}\ }\textbf {\bibinfo {volume} {91}},\ \bibinfo {pages} {235442}
  (\bibinfo {year} {2015})}\BibitemShut {NoStop}%
\bibitem [{\citenamefont {Izumida}\ \emph {et~al.}(2016)\citenamefont
  {Izumida}, \citenamefont {Okuyama}, \citenamefont {Yamakage},\ and\
  \citenamefont {Saito}}]{Izumida-2016-05}%
  \BibitemOpen
  \bibfield  {author} {\bibinfo {author} {\bibfnamefont {W.}~\bibnamefont
  {Izumida}}, \bibinfo {author} {\bibfnamefont {R.}~\bibnamefont {Okuyama}},
  \bibinfo {author} {\bibfnamefont {A.}~\bibnamefont {Yamakage}}, \ and\
  \bibinfo {author} {\bibfnamefont {R.}~\bibnamefont {Saito}},\ }\bibfield
  {title} {\enquote {\bibinfo {title} {Angular momentum and topology in
  semiconducting single-wall carbon nanotubes},}\ }\href {\doibase
  10.1103/PhysRevB.93.195442} {\bibfield  {journal} {\bibinfo  {journal} {Phys.
  Rev. B}\ }\textbf {\bibinfo {volume} {93}},\ \bibinfo {pages} {195442}
  (\bibinfo {year} {2016})}\BibitemShut {NoStop}%
\bibitem [{Note1()}]{Note1}%
  \BibitemOpen
  \bibinfo {note} {Such a treatment is justified for nanoscaled objects such as
  a carbon nanotubes in which the typical energy scale of lattice deformation
  ($\sim $ meV) is much higher than that of the rotational motion.}\BibitemShut
  {Stop}%
\bibitem [{\citenamefont {Hehl}\ and\ \citenamefont {Ni}(1990)}]{Hehl1990}%
  \BibitemOpen
  \bibfield  {author} {\bibinfo {author} {\bibfnamefont {F.~W.}\ \bibnamefont
  {Hehl}}\ and\ \bibinfo {author} {\bibfnamefont {W.-T.}\ \bibnamefont {Ni}},\
  }\bibfield  {title} {\enquote {\bibinfo {title} {Inertial effects of a
  {D}irac particle},}\ }\href@noop {} {\bibfield  {journal} {\bibinfo
  {journal} {Phys. Rev. D}\ }\textbf {\bibinfo {volume} {42}},\ \bibinfo
  {pages} {2045} (\bibinfo {year} {1990})}\BibitemShut {NoStop}%
\bibitem [{\citenamefont {Matsuo}\ \emph {et~al.}(2017)\citenamefont {Matsuo},
  \citenamefont {Saitoh},\ and\ \citenamefont {Maekawa}}]{Matsuo-JPSJ-2017}%
  \BibitemOpen
  \bibfield  {author} {\bibinfo {author} {\bibfnamefont {M.}~\bibnamefont
  {Matsuo}}, \bibinfo {author} {\bibfnamefont {E.}~\bibnamefont {Saitoh}}, \
  and\ \bibinfo {author} {\bibfnamefont {S.}~\bibnamefont {Maekawa}},\
  }\bibfield  {title} {\enquote {\bibinfo {title} {Spin-mechatronics},}\ }\href
  {\doibase 10.7566/JPSJ.86.011011} {\bibfield  {journal} {\bibinfo  {journal}
  {J. Phys. Soc. Jpn}\ }\textbf {\bibinfo {volume} {86}},\ \bibinfo {pages}
  {011011} (\bibinfo {year} {2017})}\BibitemShut {NoStop}%
\bibitem [{\citenamefont {Landau}\ and\ \citenamefont
  {Lifshitz}(1977)}]{Landau-QM-1977}%
  \BibitemOpen
  \bibfield  {author} {\bibinfo {author} {\bibfnamefont {L.~D.}\ \bibnamefont
  {Landau}}\ and\ \bibinfo {author} {\bibfnamefont {E.~M.}\ \bibnamefont
  {Lifshitz}},\ }\href@noop {} {\emph {\bibinfo {title} {Quantum Mechanics
  ---Non-relativistic theory (Third Edition)}}}\ (\bibinfo  {publisher}
  {Pergamon Press},\ \bibinfo {address} {Oxford, England},\ \bibinfo {year}
  {1977})\BibitemShut {NoStop}%
\bibitem [{Sup()}]{Supplement}%
  \BibitemOpen
  \href@noop {} {}\bibinfo {howpublished} {See Supplemental Material at
  http://link.aps.org/supplemental/10.1103/PhysRevLett.***.****** for a
  detailed information on matrix elements of the Hamiltonian, the harmonic
  potential approximation, classical solutions, derivation of the master
  equation, torque at higher temperature, the estimate for the angular
  velocity, and effect of asymmetric perturbation.}\BibitemShut {Stop}%
\bibitem [{\citenamefont {Biedenharn}\ and\ \citenamefont
  {Louck}(1984)}]{Biedenharn1984}%
  \BibitemOpen
  \bibfield  {author} {\bibinfo {author} {\bibfnamefont {L.~C.}\ \bibnamefont
  {Biedenharn}}\ and\ \bibinfo {author} {\bibfnamefont {James~D.}\ \bibnamefont
  {Louck}},\ }\href {\doibase DOI: 10.1017/CBO9780511759888} {\emph {\bibinfo
  {title} {Angular Momentum in Quantum Physics: Theory and Application}}}\
  (\bibinfo  {publisher} {Cambridge University Press},\ \bibinfo {address}
  {Cambridge},\ \bibinfo {year} {1984})\BibitemShut {NoStop}%
\bibitem [{Note2()}]{Note2}%
  \BibitemOpen
  \bibinfo {note} {The higher branches that are not shown in Fig.~\ref
  {fig:en-k} include the nutation state in which $\theta $ fluctuates around an
  average angel $\theta _0 = \protect \sqrt { \hbar |M-k| / I_\xi \omega }$.
  See Supplental Material for a detail.}\BibitemShut {Stop}%
\bibitem [{Note3()}]{Note3}%
  \BibitemOpen
  \bibinfo {note} {The recent quantum transport experiments with carefully
  fabricated carbon nanotube hybrid structures reveal the function of quantum
  dots~\cite {RevModPhys.87.703}. In general, quantum levels in a nanotube are
  highly sensitively affected by contacted electrodes~\cite
  {Izumida-2015-06,Izumida-2016-05}. However, we expect our rotor is an ideal
  carbon nanotube quantum dot since our proposed setup of the experiment is
  free from the contacts of the electrodes.}\BibitemShut {Stop}%
\bibitem [{\citenamefont {Servantie}\ and\ \citenamefont
  {Gaspard}(2006)}]{Servantie2006}%
  \BibitemOpen
  \bibfield  {author} {\bibinfo {author} {\bibfnamefont {J.}~\bibnamefont
  {Servantie}}\ and\ \bibinfo {author} {\bibfnamefont {P.}~\bibnamefont
  {Gaspard}},\ }\bibfield  {title} {\enquote {\bibinfo {title} {Translational
  dynamics and friction in double-walled carbon nanotubes},}\ }\href {\doibase
  10.1103/PhysRevB.73.125428} {\bibfield  {journal} {\bibinfo  {journal} {Phys.
  Rev. B}\ }\textbf {\bibinfo {volume} {73}},\ \bibinfo {pages} {125428}
  (\bibinfo {year} {2006})}\BibitemShut {NoStop}%
\bibitem [{\citenamefont {Sun}\ \emph {et~al.}(2005)\citenamefont {Sun},
  \citenamefont {Sato}, \citenamefont {Suenaga}, \citenamefont {Okazaki},
  \citenamefont {Kishi}, \citenamefont {Sugai}, \citenamefont {Bandow},
  \citenamefont {Iijima},\ and\ \citenamefont {Shinohara}}]{Sun-2005}%
  \BibitemOpen
  \bibfield  {author} {\bibinfo {author} {\bibfnamefont {B.-Y.}\ \bibnamefont
  {Sun}}, \bibinfo {author} {\bibfnamefont {Y.}~\bibnamefont {Sato}}, \bibinfo
  {author} {\bibfnamefont {K.}~\bibnamefont {Suenaga}}, \bibinfo {author}
  {\bibfnamefont {T.}~\bibnamefont {Okazaki}}, \bibinfo {author} {\bibfnamefont
  {N.}~\bibnamefont {Kishi}}, \bibinfo {author} {\bibfnamefont
  {T.}~\bibnamefont {Sugai}}, \bibinfo {author} {\bibfnamefont
  {S.}~\bibnamefont {Bandow}}, \bibinfo {author} {\bibfnamefont
  {S.}~\bibnamefont {Iijima}}, \ and\ \bibinfo {author} {\bibfnamefont
  {H.}~\bibnamefont {Shinohara}},\ }\bibfield  {title} {\enquote {\bibinfo
  {title} {Entrapping of exohedral metallofullerenes in carbon nanotubes:
  ({C}s{C}$_{60}$)$_n${\char64}{SWNT} nano-peapods},}\ }\href {\doibase
  https://doi.org/10.1021/ja056238a} {\bibfield  {journal} {\bibinfo  {journal}
  {J. Am. Chem. Soc.}\ }\textbf {\bibinfo {volume} {127}},\ \bibinfo {pages}
  {17972} (\bibinfo {year} {2005})}\BibitemShut {NoStop}%
\end{thebibliography}
\end{document}